\documentclass
[aps,preprint,showkeys,a4paper,tightenlines,superscriptaddress]{revtex4}%
\usepackage{eurosym}
\usepackage{amsfonts}
\usepackage{amsmath}
\usepackage{amssymb}
\usepackage{graphicx}
\usepackage{subfig}
\usepackage{caption}%
\providecommand{\U}[1]{\protect\rule{.1in}{.1in}}
\providecommand{\U}[1]{\protect\rule{.1in}{.1in}}
\newtheorem{theorem}{Theorem}
\newtheorem{acknowledgement}[theorem]{Acknowledgement}

\begin{document}
\title{Nonlinear viscoelastic isolation for seismic vibration mitigation}
\author{N. Menga}
\email{[Corresponding author. ]Email: nicola.menga@poliba.it, phone number: +39 080 5962746}
\affiliation{Department of Mechanics, Mathematics and Management, Politecnico of Bari, V.le
Japigia, 182, 70126, Bari, Italy}
\affiliation{Imperial College London, Department of Mechanical Engineering, Exhibition
Road, London SW7 2AZ}
\author{F. Bottiglione}
\affiliation{Department of Mechanics, Mathematics and Management, Politecnico of Bari, V.le
Japigia, 182, 70126, Bari, Italy}
\author{G. Carbone}
\affiliation{Department of Mechanics, Mathematics and Management, Politecnico of Bari, V.le
Japigia, 182, 70126, Bari, Italy}
\affiliation{Imperial College London, Department of Mechanical Engineering, Exhibition
Road, London SW7 2AZ}
\affiliation{CNR - Institute for Photonics and Nanotechnologies U.O.S. Bari, Physics
Department \textquotedblright M. Merlin\textquotedblright, via Amendola 173,
70126 Bari, Italy}
\keywords{RLRB, viscoelastic damping, nonlinear friction, base isolation, robustness}
\begin{abstract}
The aim of this paper is to assess the effectiveness of nonlinear viscoelastic
damping in controlling base-excited vibrations. Specifically, the focus is on
investigating the robustness of the nonlinear base isolation performance in
controlling the system response due to a wide set of possible excitation
spectra. The dynamic model is derived to study a simple structure whose base
isolation is provided via a Rubber-Layer Roller Bearing (RLRB) (rigid
cylinders rolling on rigid plates with highly damping rubber coatings)
equipped with a nonlinear cubic spring, thus presenting both nonlinear damping
and stiffness. We found that, under periodic loading, due to the non-monotonic
bell-shaped viscoelastic damping arising from the viscoelastic rolling
contacts, different dynamic regimes occur mostly depending on whether the
damping peak is overcome or not. Interestingly, in the former case, poorly
damped self-excited vibrations may be triggered by the steep damping decrease.

Moreover, in order to investigate the robustness of the isolation performance,
we consider a set of real seismic excitations, showing that tuned nonlinear
RLRB provide loads isolation in a wider range of excitation spectra, compared
to generic linear isolators. This is peculiarly suited for applications (such
as seismic and failure engineering) in which the specific excitation spectrum
is unknown a priori, and blind design on statistical data has to be employed.

\end{abstract}
\maketitle

\section{Introduction}

Controlling structural vibration is a long-standing problem and a key design
requirement in several applications, as witnessed by the huge effort made in
the last decades in developing innovative devices, and associated modelling
methodology, specifically suited for this purpose. Moreover, the range of
engineering branches facing this problem is definitely wide as all systems
involving moving parts lead to vibrations.

It is, for instance, the case of wind turbine whose vibration generates both
from internal (e.g. rotordynamics balancing, powertrain transmission) and
external (wind action, marine waves) sources. In this case, several
controlling techniques have been proposed \cite{Rahman2015}, mostly based on
passive mass-dampers \cite{Zuo2017} in the form of both ball vibration
absorbers \cite{Chen2013,Jie2012} and tuned liquid column
\cite{Colwell2009,Zhang2016}. Active techniques have also been tested using,
for instance, synthetic jets to reduce aerodynamically induced blade
vibrations \cite{Maldonado2009}. Similarly, also the isolation of large
industrial machineries is a challenging issue \cite{Rivin1995}. In this case,
regardless of the vibration source, specific solutions have been developed to
fulfill NVH requirements, such as adopting hybrid spring-actuator systems
\cite{Daley2008}, semi-active pneumatic suspension with tunable inflating
pressure \cite{Nieto2010}, and active actuators with specific control
strategies \cite{Farshidianfar2012}.

Of course, active systems are usually able to provide higher performance in
vibration control; however, these are often costly while offering relatively
low reliability due to their intrinsic higher complexity. Furthermore, their
applicability to large scale problems is usually limited by the availability
of appropriate actuators. Passive systems are therefore preferred when dealing
with sufficiently large systems. It is the case, for instance, of ensuring
high reliability to primary building such as hospitals, schools, museums and
power plants \cite{Myslimaj2003}, as well as structural control in bridges,
with strong social, political, and economic implications. In this view,
nonlinear passive isolation techniques are probably the most promising, as the
nonlinearity can lead to better isolation performance in a much wider input
spectrum. For these reasons, several possible source of nonlinearity have been
explored in the last few decades. Indeed, the piecewise nonlinear behavior
arising in both frictional dissipators \cite{DelaCruz2007} and bi-component
sacrificial connections \cite{Foti2013} has been exploited aiming at
controlling the inter-story drift caused by earthquake excitations in
high-rise structures, showing good results in terms of reduction of peak
loads. On the contrary, for low-rise heavy buildings base isolation is usually
preferred. Among them, several rolling isolation systems have been proposed
(see Ref. \cite{Harvey2016}), most of which, by employing non-planar rolling
paths, are able to produce a high-order nonlinear re-centering force
\cite{Harvey2015,Casey2018}.

Similarly, vibration mitigation also by means of nonlinear damping has been
increasingly explored \cite{Starosvetsky2009,Gendelman2015}, mostly because of
their ability to overcome typical limitations of linear dampers
\cite{Pazooki2018} (e.g. large reaction forces at high velocity, narrow
effective frequency bandwidth) when dealing, for instance, with wind or
earthquake nonstationary stochastic excitations. Among the others, these
include nonlinear energy sink (NESs), particle impact dampers (PIDs), and
nonlinear viscous dampers (NVDs), each of which presents his own specific
advantage with respect to linear systems, as indeed reviewed in Ref.
\cite{Lu2018}. Although, NESs and PIDs have been successfully applied in
controlling civil structures response to seismic motion
\cite{Nucera2008,Wierschem2017,Wang2015,Tian2017,Nakamura2009}, their primary
fields of application range from aerospace
\cite{Lee2007a,Lee2007b,Kushida2013,Ahmad2017} to machinery
\cite{Gourc2013,Gourc2015,Ema2000,Xiao2016} and lifeline
\cite{Samani2012,Tian2015,Fu2017,Jo1989,Bukhari2018} engineering. On the
contrary,\ NVDs have been increasingly utilized in civil engineering in the
recent years \cite{Symans2008,Constantinou1993,Lee2001}. This is because their
damping force only depends on velocity, thus it results out-of-phase with the
structural deformation and eventually leads to an overall reduction of loads
and displacements on the system \cite{Lu2008}. Furthermore, the damping
nonlinearity may lead to wider hysteretic cycles, associated to larger energy
dissipation \cite{Lin2008,menga2017}.

Most of NVDs rely on damping force curves in the form $F_{d}=C$sgn$\left(
\dot{x}\right)  \left\vert \dot{x}\right\vert ^{\alpha}$ which, regardless of
the $\alpha$ exponent value, leads to a monotonic response with the
deformation velocity $\dot{x}$. However, NVDs based on viscoelastic materials
result in both nonlinear and non-monotonic damping behavior \cite{menga2017},
thus leading to a limit on the maximum damping force, with possible benefits
on the overall load transmissibility. NVDs devices for base isolation of civil
structures from earthquake shocks have been already proposed so far in the
form of Rubber-Layer Roller Bearings\ (RLRB), where both metal balls
\cite{Foti1996,Foti1996b,Muhr1997,MuhrConf,Guerreiro2007} and rollers
\cite{Foti2013b,patentFoti2015,Foti2019} are interposed between plates coated
with viscoelastic rubber layers. Although very effective in dissipating
horizontal seismic components, these systems my suffer uplifts due to severe
vertical accelerations in near-fault earthquake \cite{Mazza2018}, which can
only be partially remotely accommodated into the deformable rubber coatings.
Nonetheless, the investigation of the base isolation performance of an RLRB
viscoelastic damper with linear stiffness has recently shown promising results
\cite{menga2019rlrb}, as significantly lower stresses can be achieved on a
simple base-isolated structure, in comparison with linear base isolation, as a
result of the optimization of the nonlinear damping on a specific seismic
excitation. However, since real process are always stochastic at some degree
(e.g earthquakes, machinery failure, wind) \cite{Shinozuka1988}, excitation
spectra can be unknown and blind procedures have to be employed to design
specific isolation systems, which are therefore required to present the
broadest possible effective frequency bandwidth. For these reasons, here we
focus on a fully nonlinear base isolation scheme relying on nonlinear
viscoelastic damping (through an RLRB) associated with nonlinear stiffness
(via a cubic spring). The aim is to asses whether full nonlinearity can
effectively increase the isolation performance robustness over a set of
excitations with different spectra, such as different seismic shocks. To
pursue this target, building on an accurate previously developed
\cite{menga2016visco,menga2016,menga2019coated,menga2019belt} Boundary Element
Method (BEM) description of the contact interaction between the rigid rollers
and the viscoelastic thin coating, we firstly investigate the nonlinear
dynamics of such a system under periodic excitation, also investigating
self-excited vibration induced by the steep decreasing portion of the
bell-shaped viscoelastic damping force; then, assessing the performance
robustness, we globally optimize both the nonlinear RLRB and the generic
linear isolator over a set of five real seismic shocks, eventually comparing
the performance of the two systems for each single shock in the globally
optimized conditions.

\section{Formulation}

\begin{figure}[ptbh]
\centering\includegraphics[width=0.7\textwidth]{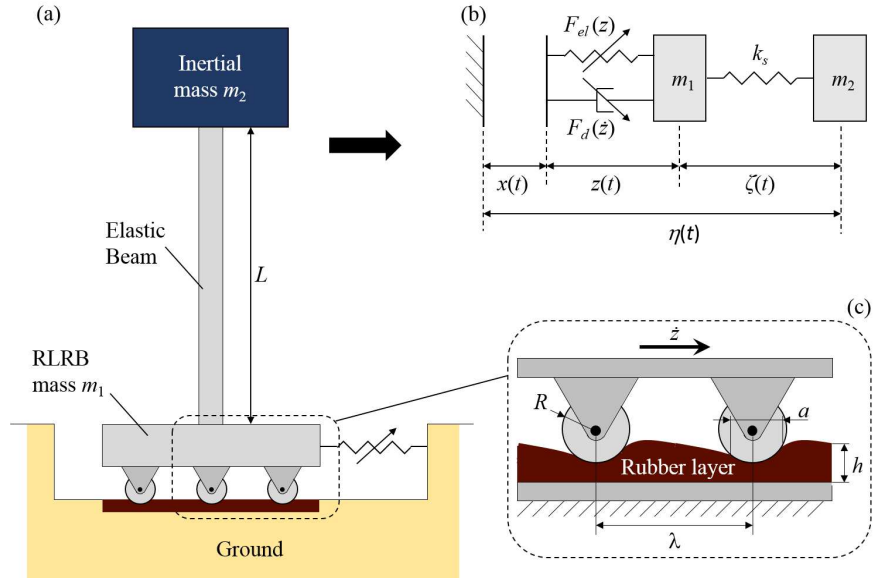}\caption{A sketch of
the two degree-of-freedom base-isolation scheme under investigation (a). The
ground vibration $x$ is filtered by means of the RLRB nonlinear damping force
$F_{d}$, and the nonlinear elastic term $F_{el}$. The inertial mass $m_{2}$ is
supported by an elastic beam of linear stiffness $k_{s}$. On the top right
(b): a lumped element scheme where $z$ and $\zeta$ are the relative
displacement between the ground and the RLRB, and RLRB and inertial mass,
respectively. On the bottom right (c): an RLRB close-up showing the rolling
contact between the rigid rollers of radius $R$ and the viscoelastic rubber
layer of thickness $h$. Due to viscoelasticity, asymmetric contact occurs,
thus leading to the nonlinear damping force $F_{d}$.}%
\label{fig1}%
\end{figure}

A functional scheme of the system under investigation is shown in Figure
\ref{fig1}a where a RLRB device of mass $m_{1}$ equipped with a nonlinear
spring is interposed between the ground and a very simple superstructure
represented by an heavy inertial mass $m_{2}$ supported by an elastic beam
with bending stiffness $k_{s}$. Since in this study we aim at investigating
the fundamental dynamic behavior of the system, we restrict our study to the
case of purely horizontal excitations.

As shown in the lumped scheme reported in Figure \ref{fig1}b, two parallel
nonlinear elements connect the superstructure to the ground: the RLRB
nonlinear damper, and the nonlinear elastic element. Regarding the latter, in
this study we focus on a cubic nonlinear spring, so that $F_{el}\left(
z\right)  =\mu z^{3}$, where $z$ is the RLRB displacement, i.e. the relative
displacements between the ground and the superstructure base.

Within the RLRB, a nonlinear damping force arises due to the rolling contact
between the rigid rollers and the viscoelastic coating, as already shown in
Refs. \cite{menga2017,menga2019rlrb}. Specifically, the viscoelastic material
response is represented by the viscoelastic complex modulus $E\left(
\omega\right)  $, which is a nonlinear function of the excitation frequency
$\omega$. In the case of a single relaxation time $\tau$ material, we have that%

\begin{equation}
E\left(  \omega\right)  =E_{0}+E_{1}\frac{\mathrm{i}\omega\tau}{1+\mathrm{i}%
\omega\tau} \label{Eomega}%
\end{equation}
where $E_{1}=E_{\infty}-E_{0}$, with $E_{0}$ and $E_{\infty}$ being the zero
and high frequency elastic moduli, respectively. In rolling contacts such as
that occurring in the RLRB between the rollers and the viscoelastic coating
(see Figure \ref{fig1}c), the relative motion results in a cyclic deformation
of the viscoelastic layer which, due to the viscoelastic response delay, may
lead to asymmetric contact pressure distributions and, in turn, to a damping
force $F_{d}\propto Im[E(\omega)]$ opposing the relative motion
\cite{christensen,menga2016visco}. Notably, since $\omega\approx\dot
{z}/\lambda$, we have that $F_{d}=F_{d}\left(  \dot{z},\tau,\lambda\right)  $.
Furthermore, from Eq. (\ref{Eomega}) we have that $Im[E(\omega)]$ is maximum
for $\omega\tau\approx1$, thus the maximum damping force can be reasonably
expected for $\dot{z}^{\ast}\approx\lambda/\tau$. Similarly, it can be shown
that for very low and very high values of $\dot{z}$, the damping force
vanishes, as $Im[E(\omega\rightarrow0)]=Im[E(\omega\rightarrow\infty)]=0$.
However, in order to calculate the exact viscoelastic damping force as a
function of $\dot{z}$, the contact problem between the rollers and the rubber
layer has to be solved numerically, assuming both $\lambda$ and $\tau$ as a
parameter. This can be done by relying on the boundary element formulation
given in Refs. \cite{menga2018visco,menga2019coupling}, where the specific
viscoelastic layer thickness $h$ and roller radius $R$ are also taken into account.

\begin{figure}[ptbh]
\centering\includegraphics[width=0.5\textwidth]{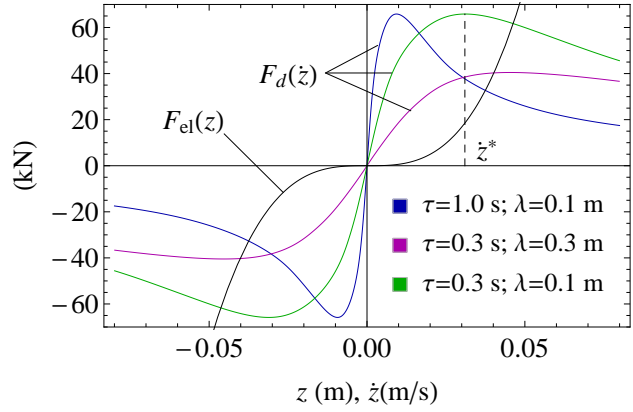}\caption{The
nonlinear elastic force $F_{el}$ and damping force $F_{d}$ as functions of the
ground-RLRB relative displacement $z$ and velocity $\dot{z}$, respectively.
The damping force $F_{d}$ is shown for different values of the viscoelastic
material relaxation time $\tau$ and contact wavelength $\lambda$. Results are
for $h/\lambda=0.13$.}%
\label{fig2}%
\end{figure}

In Figure \ref{fig2} the trend the nonlinear forces $F_{el}\left(  z\right)  $
and $F_{d}\left(  \dot{z}\right)  $ is shown. Interestingly, in the same
figure the effect of the RLRB parameters $\tau$ and $\lambda$ on $F_{d}$ can
be appreciated: as indicated above, the ratio $\tau/\lambda$ shifts the value
of the peak velocity $\dot{z}^{\ast}$; notably, $\lambda$ also affects the
peak amplitude. This is mainly due to the influence of $\lambda$ on the mean
contact pressure $\bar{p}$ (i.e. on the amount of viscoelastic material
cyclically deformed) which is given by the vertical static equilibrium of the
rollers as%
\begin{equation}
\bar{p}=g\frac{m_{1}+m_{2}}{N\lambda b}%
\end{equation}
where $g$ is the gravitational acceleration, $b$ is the RLRB transverse width,
and $N$ is the number of rollers.

Finally, referring to the lumped scheme of Figure \ref{fig1}b, the linear
momentum balance gives
\begin{equation}%
\begin{array}
[c]{c}%
m_{1}\left(  \ddot{x}+\ddot{z}\right)  +\mu z^{3}+F_{d}\left(  \dot{z}\right)
-k_{s}\zeta=0\\
m_{2}\left(  \ddot{x}+\ddot{z}+\ddot{\zeta}\right)  +k_{s}\zeta=0
\end{array}
\label{eqsys}%
\end{equation}
with
\begin{equation}
\eta(t)=x(t)+z(t)+\zeta(t).
\end{equation}

In order to integrate the set of nonlinear second order ODE of Eqs.
(\ref{eqsys}), we adopted a fixed-step numerical strategy based on fourth
order \textit{Runge-Kutta }integration algorithm \cite{Demidovich}. The choice
of the integration time-step has followed a sensitivity analysis aiming at
avoiding numerical instabilities.

Aiming at comparing the nonlinear system results in terms of dynamic response
under periodic base excitation, a Duffing-like version of the system at hand
(i.e. $F_{d}=c_{eq}\dot{z}$ in Eq. (\ref{eqsys})) is investigated by relying
on the approximate \textit{method of iterations} \cite{Thomson} (see Appendix
A). Indeed, in agreement with Ref. \cite{menga2019rlrb}, for each set of
viscoelastic parameters ($\tau$, $\lambda$) a different damping curve exists
(see Figure \ref{fig2}), and an equivalent linear damping coefficient $c_{eq}$
can be defined as
\begin{equation}
c_{eq}=\left.  -\frac{\mathrm{d}F_{d}}{\mathrm{d}\dot{z}}\right\vert _{\dot
{z}=0}%
\end{equation}

Finally, it is noteworthy to observe that, according to Refs.
\cite{menga2017,menga2018visco}, viscoelastic calculations have been performed
assuming steady rolling (i.e. constant velocity) between the RLRB rollers and
the rubber layer, whereas periodic/seismic ground oscillation would reasonably
induce reciprocating relative motion conditions. Nonetheless, as reported in
Ref. \cite{menga2019rlrb}, for $a\ll s$\emph{ }and\emph{ }$\tau\ll T$, with
$s$ and $T$ being the stroke and period of the external forcing, steady
sliding calculations can still provide a physically meaningful approximation
of the real viscoelastic response.

\section{Results}

Calculations have been performed assuming an incompressible (i.e. $\nu=0.5$)
viscoelastic material with a single relaxation time and $E_{0}=E_{\infty
}/3=50$ MPa. Furthermore, in terms of RLRB geometry, we assumed $h/\lambda
=0.13$, $R/\lambda=0.3$, and $b=1$ m.

Since RLRBs are usually employed in the isolation of buildings such as
hospitals, schools and museums \cite{Myslimaj2003}, we assumed the dynamic
model parameters in order to achieve a qualitative overlap with these possible
applications. Indeed, according to Ref. \cite{menga2019rlrb}, we set
$m_{1}=1\times10^{2}$ kg, $m_{2}=1\times10^{5}$ kg. The supporting beam
stiffness has been estimated as the bending stiffness of a commercial HEB 300
steel beam, with $L=3$ m, thus resulting in $k_{s}=6\times10^{6}$ N/m.
Finally, to simplify the results reading, the nonlinear stiffness will be
written as $\mu=\mu_{0}6.2\times10^{7}$N/m$^{3}$, with $\mu_{0}$ being a
dimensionless coefficient.

\subsection{Periodic dynamics}

In this section we focus on the dynamic response of the physical system under
periodic base excitation in the form $x\left(  t\right)  =x_{0}\sin\left(
\omega t\right)  $. Since the system is nonlinear both in terms of stiffness
and damping, we expect the dynamic behavior to be affected by the initial
conditions; therefore, we numerically simulate both forward and backward
excitation frequency sweep, for different values of both the excitation
amplitude $x_{0}$ and damping conditions.

\begin{figure}[ptbh]
\centering\includegraphics[width=0.95\textwidth]{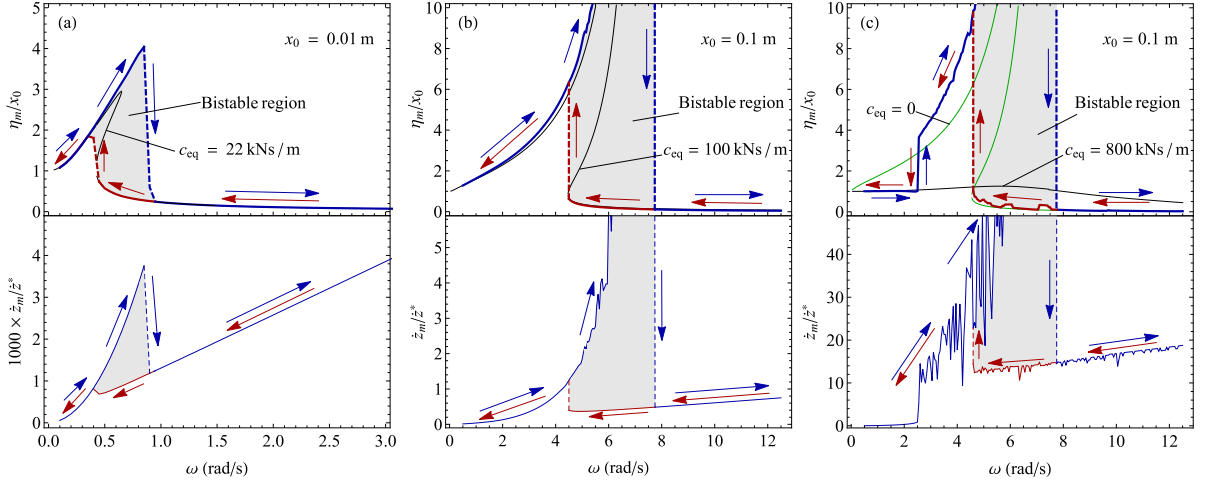}\caption{The
normalized amplitude of the inertial mass oscillation $\eta_{m}/x_{0}$ (upper
diagrams) and the relative velocity amplitute $\dot{z}_{m}/\dot{z}^{\ast}$
between the ground and RLRB (lower diagrams) as a function of the excitation
frequency $\omega$. Blue and red curves represent forward and backward
frequency sweep stable results for the nonlinear system, respectively. Black
curves represent equivalent Duffing-like approximate results, whereas the
green curve is the undamped approximate result. The dimensionless nonlinear
stiffness is $\mu_{0}=1$, and the nonlinear damping parameters are
$\lambda=0.2$ m and (a) $\tau=0.002$ s, (b) $\tau=0.0095$ s, (c) $\tau=0.0742$
s.}%
\label{fig3}%
\end{figure}

The frequency response is shown in Figures \ref{fig3} in terms of the
amplitude of both the inertial mass oscillation $\eta_{m}$ (upper diagrams)
and the relative velocity between the ground and RLRB $\dot{z}_{m}$ (lower
diagrams). Specifically, Figure \ref{fig3}a refers to a sufficiently small
value of $x_{0}$. The equivalent Duffing-like system with linear damping
(black curve) shows the well-known stiffening multivalued frequency response
(black curve), which overhangs to the high-frequency side, thus leading to a
bistable region. Under these conditions, since $\dot{z}_{m}/\dot{z}^{\ast}<1$,
the system equipped with the RLRB is poorly affected by the damping
nonlinearity, thus a slightly less damped response is observed (see blue and
red curves in Figure \ref{fig3}a) associated to an overall larger oscillation
amplitude compared to the Duffing-like system. Although in Figure \ref{fig3}b
different parameters for damping and excitation amplitude are investigated,
the physical scenario described above still qualitatively applies.
Nonetheless, this time significantly larger values of $\eta$ are expected, as
forward swapping results show fast increasing oscillation amplitudes,
eventually approaching the vertical asymptote at $\omega_{1}\approx7.75$ rad/s
(see Appendix A). Beyond this threshold, single-valued results are recovered,
with forward and backward stable points overlapping. A similar behavior is
observed in the lower diagram of Figure \ref{fig3}b concerning the amplitude
of $\dot{z}$. In Figure \ref{fig3}c a dramatically different system response
is observed. Indeed, contrary to the cases discussed above, this time
significant differences are reported between the Duffing-like system behavior
and the RLRB. Specifically, this time, the choice of $\tau$ and $\lambda$
\ leads to equivalent larger damping (i.e. lower value of $\dot{z}^{\ast}$)
entailing more severe peak velocity overcoming, as indicated by the lower
diagram where $\dot{z}_{m}/\dot{z}^{\ast}\gg1$. Under these conditions, the
nonlinear system is able to replicate the highly damped Duffing-like results
only at low excitation frequency, whereas the qualitative agreement is lost at
larger $\omega$. In the same figure, the undamped system behavior is reported
for comparison (i.e. $F_{d}=0$ in Eq. (\ref{eqsys})). Focusing on the upper
overhanging branch, the undamped system present even lower amplitudes than the
nonlinearly damped one. Such a result is peculiar of system presenting
portions of the damping curve decreasing with increasing velocity such as
those based on viscoelastic friction. In this case, indeed, the abrupt
decrease of the damping force for $\left\vert \dot{z}\left(  t\right)
\right\vert >\dot{z}^{\ast}$ generates additional self-excited components to
the system vibrational response. Notably, the occurrence of such a phenomenon
is mostly governed by the sharpness of the damping force decrease beyond the
peak, thus resulting noteworthy in our system only at sufficiently large
values of $c_{eq}$ (see Figure \ref{fig2}).

\begin{figure}[ptbh]
\centering\subfloat[\label{fig4a}]{\includegraphics[height=5.2cm]{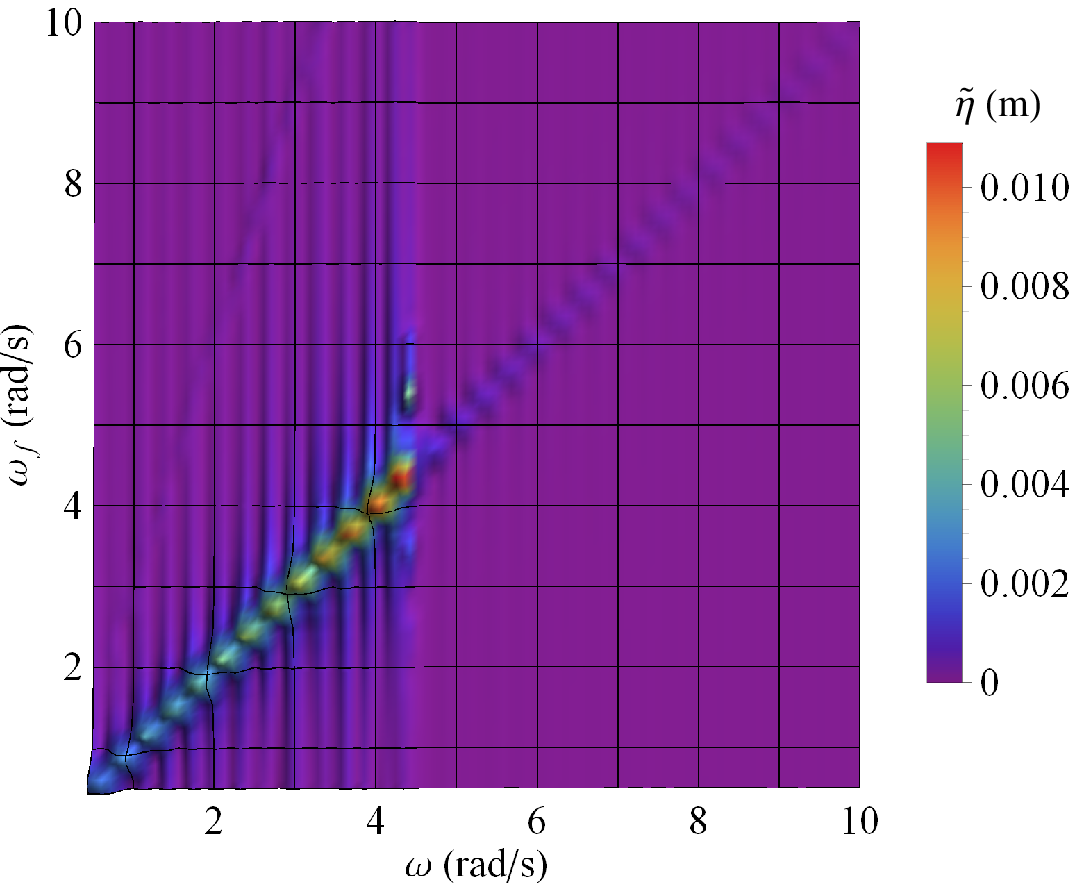}}
\hspace{1 cm}
\subfloat[\label{fig4b}]{\includegraphics[height=5.2cm]{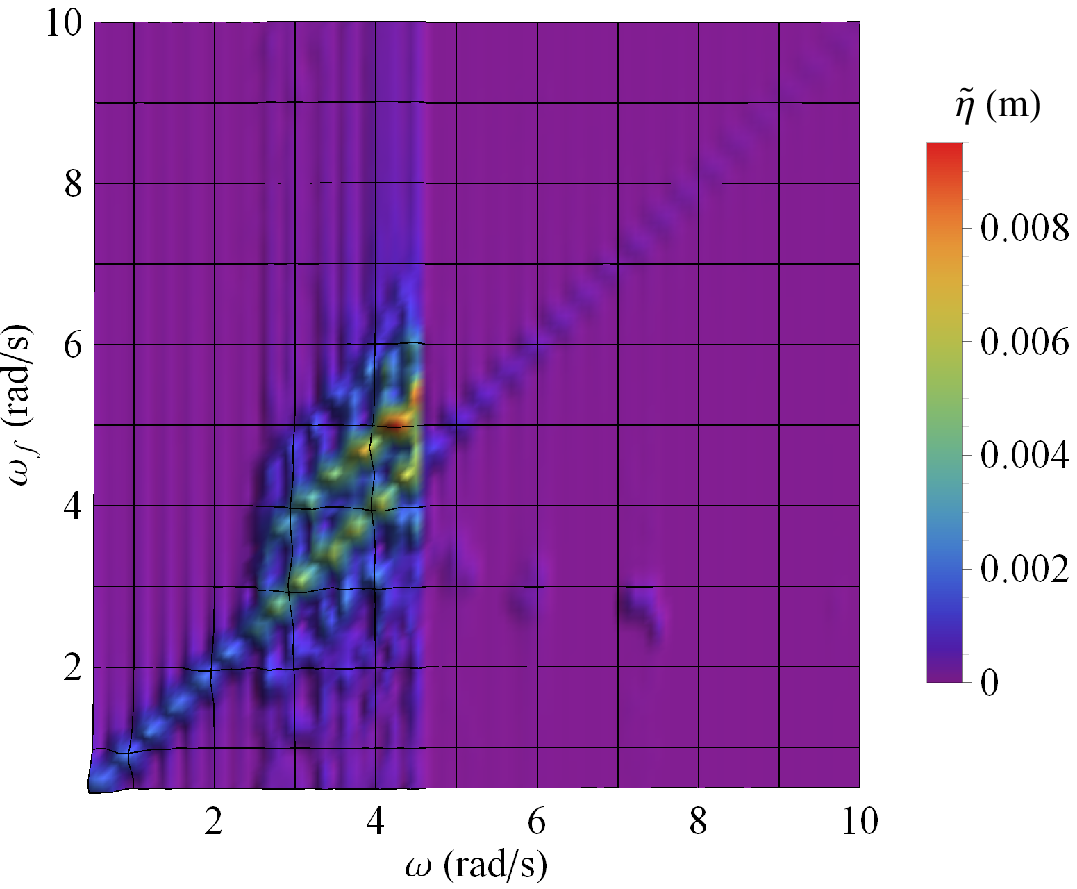}}\caption{The
spectrograms of the oscillation amplitude $\tilde{\eta}$ vs the peridic
excitation frequency $\omega$ for the backward stable branches of (a) Figure
\ref{fig3}b, and (b) Figure \ref{fig3}c. Results are for $x_{0}=0.1$ m,
$\mu_{0}=1$, $\lambda=0.2$ m and (a) $\tau=0.0095$ s, (b) $\tau=0.0742$ s.}%
\label{fig4}%
\end{figure}

Such a result is also confirmed by Figures \ref{fig4} where the spectrogram of
the oscillation amplitude $\eta$ is shown. Specifically, Figure \ref{fig4a}
refers to the backward stable branch of Figure \ref{fig3}b. As expected, in
this case the main spectral contribution is associated to the external
excitation frequency, whereas different spectral components are negligible
except in the proximity of the solution jump up ($\omega\approx4.5$ rad/s).
Similarly, Figure \ref{fig4b} refers to the backward stable branch of Figure
\ref{fig3}c. Interestingly, a significant spread of the spectrum is associated
to the peak velocity overcoming, consistently with the activation of
additional self-excited vibration components.

\begin{figure}[ptbh]
\centering\includegraphics[width=0.85\textwidth]{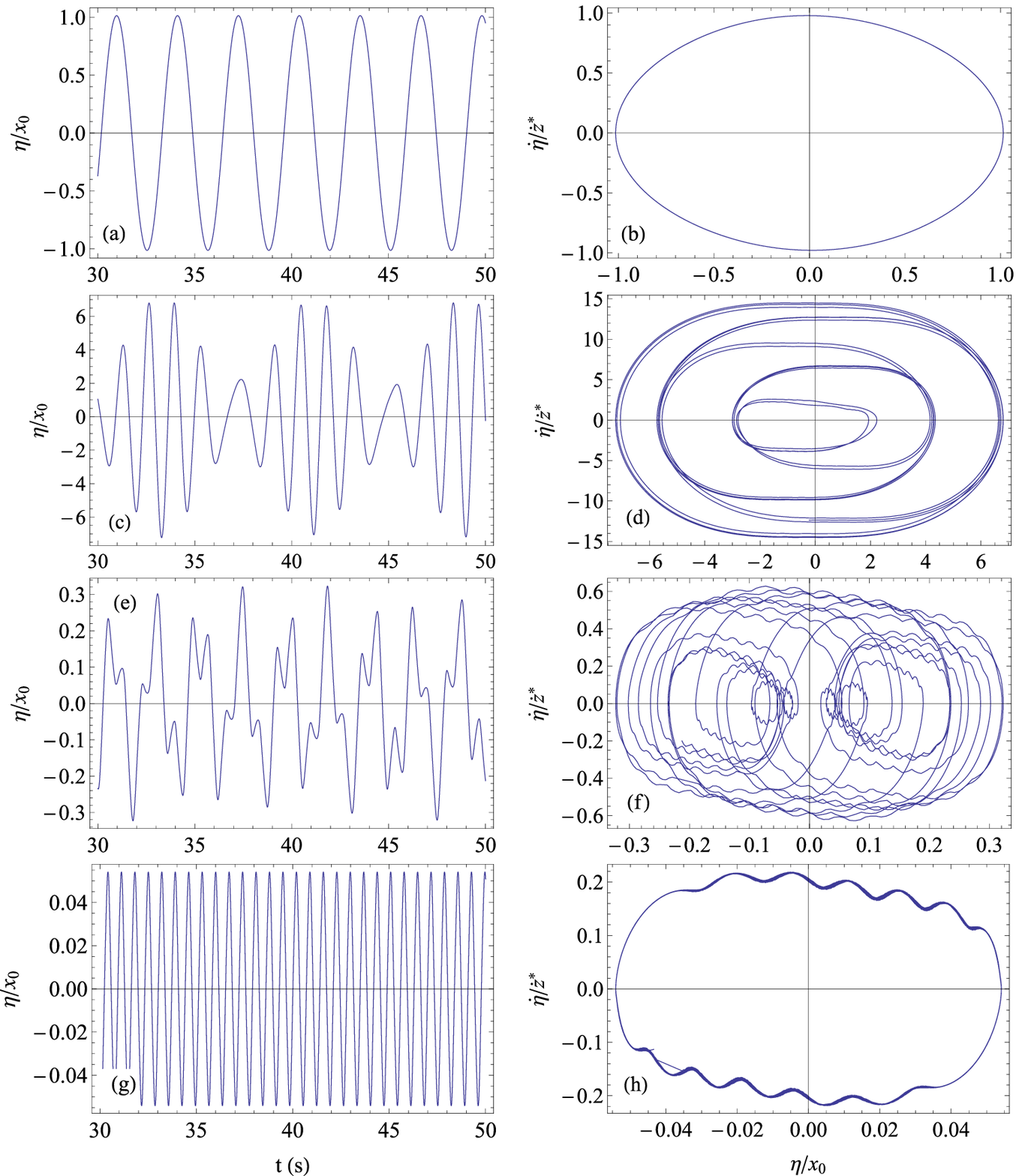}\caption{The
time history of the normalized displacement $\eta/x_{0}$ (left column), and
the normalized phase-diagram (right column). Data refer to the backward stable
branch of Figure \ref{fig3}c, with $\omega=2$ rad/s (a-b); $4$ rad/s (c-d);
$7.2$ rad/s (e-f); $9$ rad/s (g-h).}%
\label{fig4bis}%
\end{figure}

Figures \ref{fig4bis} show the inertial mass oscillation time history (left
column) and phase portrait (right column) at different values of $\omega$ on
the backward branch of Figure \ref{fig3}c. Figures \ref{fig4bis}a-b and
\ref{fig4bis}g-h share the same solution with the equivalent Duffing-like
system with linear damping, thus showing a periodic response with elliptical
phase portrait due to the slight damping nonlinearity. On the contrary,
Figures \ref{fig4bis}c-d show different results as, this time, due to the
system nonlinearity additional frequencies are excited, as indeed shown by the
phase portrait where at least five closed paths can be observed. Finally, the
responses shown in Figures \ref{fig4bis}e-f is even more complex, involving an
larger number of spectral components.

\subsection{Seismic dynamics}

The fundamental investigation performed in the previous section allows to
conclude that the nonlinear isolator under investigation can exhibit
significantly different dynamic behaviors depending on the value of the
elastic ($\mu$) and damping ($\tau,\lambda$) parameters.

In terms of seismic vibration mitigation, in Ref. \cite{menga2019rlrb}, the
performance of an RLRB equipped with linear stiffness has been compared to a
generic linear isolator on a single seismic event, showing a marked ability of
the RLRB with linear stiffness to relief the loads acting on the
superstructure. However, since the possible excitation spectrum in real
applications (e.g. earthquakes, wind, failures) is often unknown, we expect
the robustness of the isolation performance to play a key role in practical
problems. Therefore, focusing on the RLRB with nonlinear damping and stiffness
and the generic linear isolator, we perform a global optimization of both the
system assuming a series of five different seismic events as inputs; then, we
compare the specific isolation performance achieved by both systems in global
optimum conditions on each specific shock. To this regard, the linear momentum
balance for the generic linear system can be achieved as the linearized
version of Eqs. (\ref{eqsys}), so%
\begin{equation}%
\begin{array}
[c]{c}%
m_{1}\left(  \ddot{x}+\ddot{z}_{l}\right)  +k_{l}z_{l}+c_{l}\dot{z}_{l}%
-k_{s}\zeta_{l}=0\\
m_{2}\left(  \ddot{x}+\ddot{z}_{l}+\ddot{\zeta}_{l}\right)  +k_{s}\zeta_{l}=0
\end{array}
\label{linsys}%
\end{equation}
where the optimization parameters are $c_{l}$ and $k_{l}$.

Furthermore, since our study focus on a simple single-story superstructure,
most of the usual performance index \cite{Preumont2008,Yanik2014}, based on
inter-story drift and shear stresses, cannot be adopted. However, according to
Refs. \cite{Sadek1998,Ng2007}, for each considered shock we can define three
main source of damage: (i) the amplitude $F_{m}$ of the inertial force acting
on the mass $m_{2}$ (i.e. $F_{m}=k_{s}\zeta_{m}$ with $\zeta_{m}$ being the
amplitude of $\zeta$), associated to the superstructure instantaneous damage;
(ii) the root mean square $F_{rms}$ of the inertial force acting on the mass
$m_{2}$ (i.e. $F_{rms}=k_{s}/T\left[  \int_{T}\zeta^{2}\left(  t\right)
dt\right]  ^{1/2}$with $T$ being the shock duration), associated to the
material hysteresis and fatigue; (iii) the amplitude of the relative
displacement at the base $z_{m}$, associated to possible compatibility issues
with surrounding systems (e.g. buildings, piping, frameworks). Finally, the
performance index is defined as
\begin{equation}
\varphi=\frac{1}{n}\sum_{k=1}^{n}\left(  w_{1}\frac{F_{m,k}}{F_{m,0}}%
+w_{2}\frac{F_{rms,k}}{F_{rms,0}}+w_{3}\frac{z_{m,k}}{z_{m,0}}\right)
\label{phi}%
\end{equation}
where $k$ spans through the $n$ seismic events, and $w_{i}$, with $i=1,2,3$,
are weighting terms. In our analysis we assume $w_{i}=1/3$. Notably, the
smaller the value of $\varphi$, the better the performance is. Moreover, in
Eq. (\ref{phi}) $F_{m,0}$, $F_{rms,0}$, and $z_{m,0}$ are homogenization terms
defined as the absolute maxima of the corresponding quantities in the
investigated domain. The optimization process has been solved by means of a
brute-force search with structured sampling of the parameters domain.

\begin{figure}[ptbh]
\centering\includegraphics[width=0.8\textwidth]{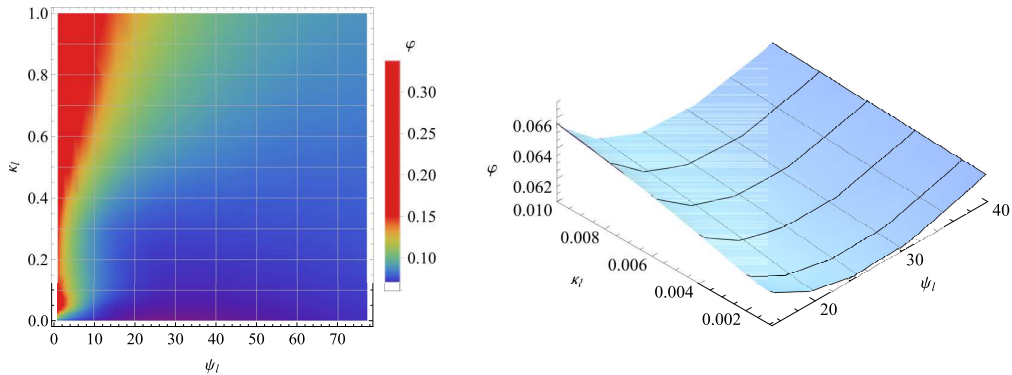}\caption{The
performance index $\varphi$ of the system equipped with generic linear
isolator as a function of the dimensionless linear stiffness $\kappa_{l}$ and
damping $\psi_{l}$ (left), with a close up to the optimal region (right). With
reference to Eqs. (\ref{linsys}), $k_{l}=\kappa_{l}\times3.1\times10^{6}$N/m
and $c_{l}=\psi_{l}\times10^{4}$Nm/s.}%
\label{fig5}%
\end{figure}

Figures \ref{fig5} show the effect of dimensionless base stiffness
($\kappa_{l}$) and damping ($\psi_{l}$) on the response of the system equipped
with a generic linear isolator. Interestingly, the base stiffness appears less
effective than the damping coefficient in determining the overall isolation
performance, thus the investigated range of stiffness has been cropped at
reasonably low values.

\begin{figure}[ptbh]
\centering\includegraphics[width=0.8\textwidth]{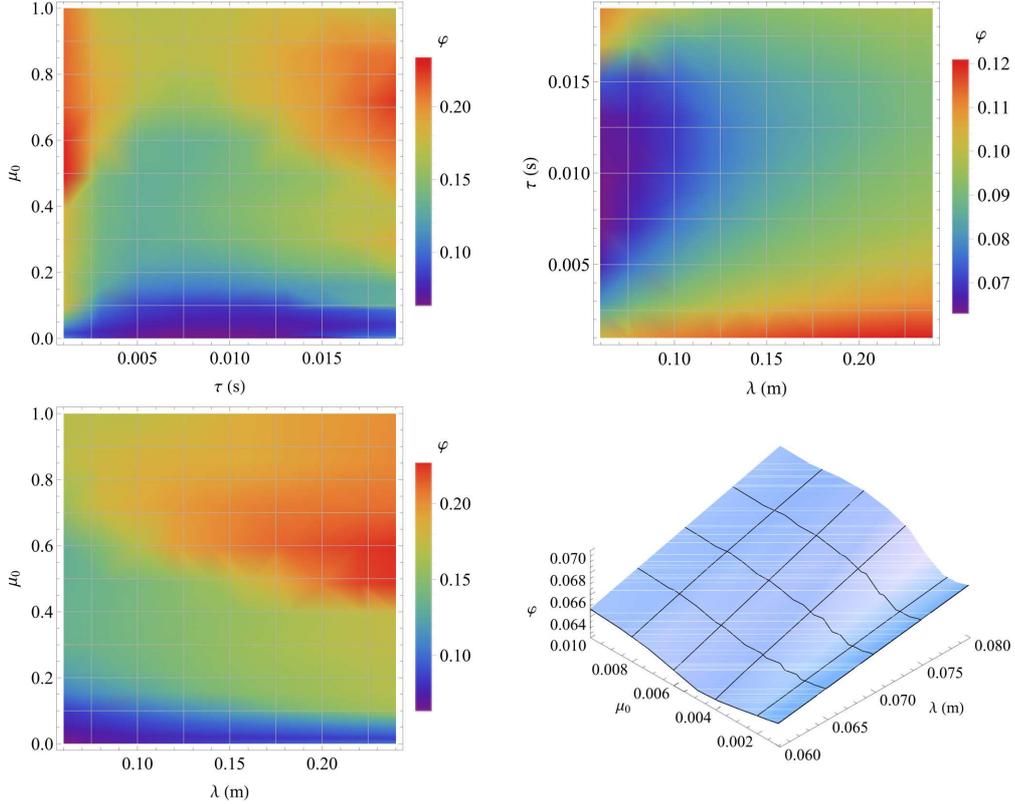}\caption{The
performance index $\varphi$ of the system equipped with nonlinear RLRB as a
function of the dimensionless nonlinear elastic $\mu_{0}$ term, and the RLRB
damping parameters $\lambda$ and $\tau$. A 3D closeup to the optimal region is
also provided.}%
\label{fig6}%
\end{figure}

Similarly, Figures \ref{fig6} show the isolation performance of the system
equipped with RLRB and nonlinear stiffness. For what concerns the damping
behavior, as expected from the discussion reported in the previous sections,
we observe a slightly larger effect of $\tau$ on the overall isolation
performance, compared to $\lambda$. The reason of such a behavior can be
ascribed to the direct effect of $\tau\,$and the slope of the damping curve,
contrary to $\lambda$ which also affects the damping force peak value. For
what concerns the nonlinear elastic term, although the optimum condition is
related to sufficiently low values of $\mu_{0}$, it is noteworthy to observe
that a non-monotonic effect on the overall isolation performance occurs.

\begin{table}[ptb]
\centering%
\begin{tabular}
[c]{ccccccc}
& \multicolumn{3}{c}{Linear} & \multicolumn{3}{c}{Nonlinear}\\
& $F_{m}^{i}$ (kN) & $F_{rms}^{i}$ (kN) & $z_{m}$ (cm) & $F_{m}^{i}$ (kN) &
$F_{rms}^{i}$ (kN) & $z_{m}~$(cm)\\
Irpinia 1980 & 92.7 & 15.3 & 19.5 & 66.3 & 13.2 & 23.9\\
Izmit 1999 & 102 & 15.5 & 28.3 & 69.4 & 14.2 & 36.7\\
Waiau 2016 & 245 & 22.9 & 16.8 & 81.1 & 16.5 & 23.9\\
Kathmandu 2015 & 103 & 20.2 & 31.2 & 100 & 21.0 & 57.7\\
Christchurch 2010 & 223 & 31.3 & 25.5 & 89.4 & 21.0 & 40.4\\
Mean & 153 & 21.0 & 24.6 & 81.2 & 17.2 & 36.5\\
Standard deviation & 74.3 & 6.58 & 6.0 & 14 & 3.69 & 14.0
\end{tabular}
\caption{Optimization results for both the nonlinear and linear system
performance. Optimal conditions are $\kappa_{l}=0.001$ and $\psi_{l}=24.59$
for the linear system, and $\mu_{0}=0.0048$, $\lambda=0.06$ m, $\tau=0.007$ s
in the nonlinear case.}%
\label{tab1}%
\end{table}

\begin{figure}[ptbh]
\centering\includegraphics[width=0.8\textwidth]{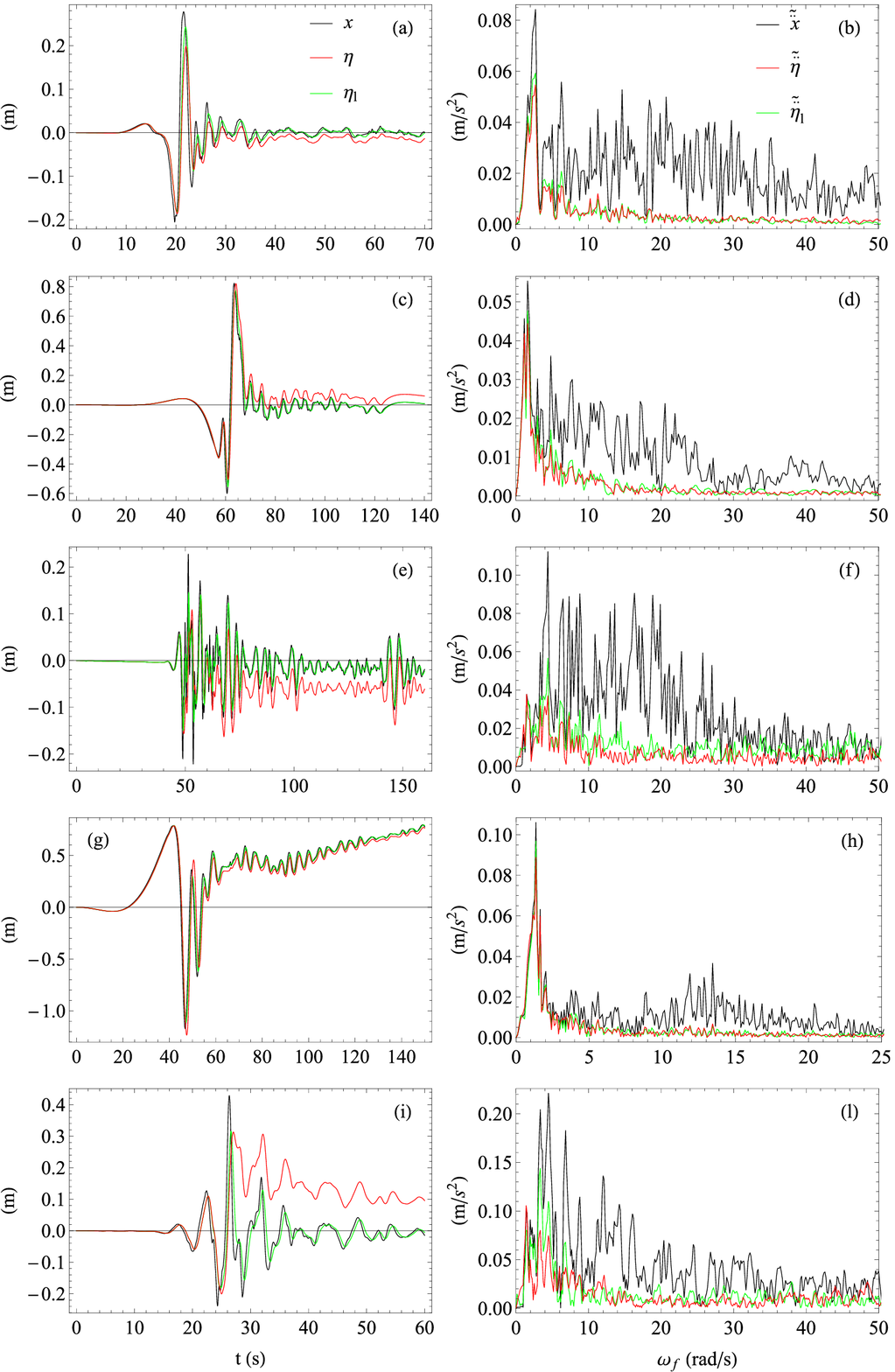}\caption{The
time history (left column) and spectral components (right column) of the
optimal responses of the system equipped with generic linear base isolation
(green curves) and nonlinear RLRB (red curves) to the considered earthquake
shocks: Irpinia 1980 (a-b); Izmit 1999 (c-d); Waiau 2016 (e-f); Kathmandu 2015
(g-h); Christchurch 2010 (i-l).}%
\label{fig7}%
\end{figure}

The isolation performance in global optimum conditions are reported in Table
\ref{tab1}, showing that, for each of the considered shocks, the nonlinear
RLRB is able to provide better loads isolation in a wider range of excitation
spectra (i.e. for different seismic events). Such result seems to indicate a
higher performance robustness associated to the nonlinear isolation with
respect to the excitation variability, with direct possible impact on several
engineering applications in which the forcing spectrum is unknown and blind
design of the isolation is required.

The optimal time histories of $\eta$ are shown in the left column in Figures
\ref{fig7} for all the seismic events here considered. We observe that, in
most of the cases, the nonlinear RLRB base isolation leads to overall smoother
responses compared to a generic linear base isolation, also slightly reducing
the peak displacement. Interestingly, a persisting displacement offset may be
observed after the shock between the superstructure and the ground in the case
of nonlinear RLRB. This has to be ascribed to the slowly decaying $z$ relative
rigid displacement between the superstructure base and the ground due to the
poor re-centering effect provided by the nonlinear base elastic term for
$z\ll1$. Of course, this displacement offset asymptotically vanishes in the
long term response. Similar conclusions can be drawn from the right column in
Figures \ref{fig7} showing the spectral representation of the inertial mass
acceleration $\ddot{\eta}$. As expected, both the linear and nonlinear base
isolation systems behave as low pass filters, however enhanced filtering
performances are associated to the nonlinear RLRB.

\section{Conclusions}

In this work, nonlinear viscoelastic vibrations are investigated with specific
focus on nonlinear damping behavior arising in viscoelastic rolling contacts
such as those involved in RLRB base insolation systems.

We found that, due to the bell-shaped damping curve peculiar of viscoelastic
materials, different dynamic regimes can be observed under periodic loading,
sometimes triggering self-excited vibrational components, with higher
oscillation amplitudes compared to equivalent systems with linear damping.

Furthermore, we compare the effectiveness of the RLRB equipped with cubic
spring in terms of real seismic base isolation against a generic linear
isolator. We performed global optimization on damping and stiffness parameters
in both cases assuming as inputs a set of five different seismic events.
Results highlight a better loads isolation performance of the nonlinear RLRB,
clearly showing that under the same global optimal conditions the nonlinear
base isolation is able to tolerate a wider range of excitations, thus
exhibiting a significantly more robust performance than the linear one. This
suggests that nonlinear isolation systems involving non-monotonic damping,
such as those based on viscoelastic friction, may represent a preferable
choice in applications requiring blind design of the isolation system.

\begin{acknowledgement}
This project has received funding from the European Union's Horizon 2020
research and innovation programme under the Marie Sk\l odowska-Curie grant
agreement No 845756 (N.M. Individual Fellowship). This work was partly
supported by the Italian Ministry of Education, University and Research under
the Programme \textquotedblleft Progetti di Rilevante Interesse Nazionale
(PRIN)\textquotedblright, Grant Protocol 2017948, Title: Foam Airless Spoked
Tire -- FASTire (G.C.).
\end{acknowledgement}

\appendix{}

\section{Duffing-like system approximate frequency response}

Here we report the method \cite{Thomson} we adopted to calculate the
approximated frequency response of the Duffing-like two-DoF system with
nonlinear elastic force $\mu x^{3}$ and linear (viscous) damping $c\dot{x}$.
We focus on the case with periodic ground excitation in the form
$x(t)=x_{0}\cos\omega t$. The linear momentum balance equations are%
\begin{align}
m_{1}\ddot{z}  &  =-\mu z^{3}-c\dot{z}+k_{s}\zeta+m_{1}x_{0}\omega^{2}%
\cos\omega t\nonumber\\
m_{2}\ddot{z}+m_{2}\ddot{\zeta}  &  =-k_{s}\zeta+m_{2}x_{0}\omega^{2}%
\cos\omega t \label{linmom}%
\end{align}

The system is nonautonomous as the time $t$ appears explicitly in the forcing
terms. We seek only the steady state harmonic solution by the method of
iteration. The first assumed solution is%
\begin{align}
z  &  =A_{1}\cos\omega t+A_{2}\sin\omega t\nonumber\\
\zeta &  =B_{1}\cos\omega t+B_{2}\sin\omega t \label{linmomsol}%
\end{align}

By substituting eq. (\ref{linmomsol}) into the differential equations
(\ref{linmom}), we found that%
\begin{align}
z  &  =\frac{1}{36m_{1}\omega^{2}}[36m_{1}\omega^{2}\left(  c_{1}%
+c_{2}t\right)  +9\left(  -4B_{1}k_{s}+3\mu A_{1}^{3}+3\mu A_{1}A_{2}%
^{2}+4c\omega A_{2}-4m_{1}\omega^{2}x_{0}\right)  \cos\omega t+\nonumber\\
&  \mu A_{1}\left(  A_{1}^{2}-3A_{2}^{2}\right)  \cos3\omega t+9\left(
-4B_{2}k_{s}+3\mu A_{1}^{2}A_{2}+3\mu A_{2}^{3}-4A_{1}c\omega\right)
\sin\omega t-\nonumber\\
&  \mu A_{2}\left(  A_{2}^{2}-3A_{1}^{2}\right)  \sin3\omega t] \label{sol_z}%
\end{align}

\begin{align}
\zeta &  =\frac{1}{36m_{1}m_{2}\omega^{2}}\{36m_{1}m_{2}\omega^{2}\left(
c_{3}+c_{4}t\right)  +9[4B_{1}k_{s}\left(  m_{1}+m_{2}\right)  -3\mu
m_{2}A_{1}\left(  A_{1}^{2}+A_{2}^{2}\right)  -\nonumber\\
&  4A_{2}m_{2}c\omega]\cos\omega t-\mu A_{1}m_{2}\left(  A_{1}^{2}-3A_{2}%
^{2}\right)  \cos3\omega t+9[4B_{2}k_{s}\left(  m_{1}+m_{2}\right) \nonumber\\
&  -3\mu A_{2}m_{2}\left(  A_{1}^{2}+A_{2}^{2}\right)  +4A_{1}m_{2}%
c\omega]\sin\omega t+\mu m_{2}A_{2}\left(  A_{2}^{2}-3A_{1}^{2}\right)
\sin3\omega t\} \label{sol_zita}%
\end{align}

Since we focus on the harmonic solution with period $T=2\pi/\omega$, we have
that $c_{1}=c_{2}=c_{3}=c_{4}=0$. Indeed, we only aim at providing a
qualitative estimation of the overall dynamic behavior, here we neglect the
higher harmonic terms. However, the method can easily be further iterated.

Collecting all the trigonometric terms in Eqs. (\ref{sol_z},\ref{sol_zita})
and considering Eqs. (\ref{linmomsol}), the following set of nonlinear
algebraic equations is found%

\begin{align}
A_{1}  &  =\frac{1}{4m_{1}\omega^{2}}\left[  -4B_{1}k_{s}+3\mu A_{1}^{3}+3\mu
A_{1}A_{2}^{2}+4A_{2}c\omega-4m_{1}\omega^{2}x_{0}\right] \nonumber\\
A_{2}  &  =\frac{1}{4m_{1}\omega^{2}}\left[  -4B_{2}k_{s}+3\mu A_{1}^{2}%
A_{2}+3\mu A_{2}^{3}-4A_{1}c\omega\right] \nonumber\\
B_{1}  &  =\frac{1}{4m_{1}m_{2}\omega^{2}}\left[  4B_{1}k_{s}(m_{1}%
+m_{2})-3\mu A_{1}m_{2}(A_{1}^{2}+A_{2}^{2})-4A_{2}m_{2}c\omega\right]
\nonumber\\
B_{2}  &  =\frac{1}{4m_{1}m_{2}\omega^{2}}\left[  4B_{2}k_{s}(m_{1}%
+m_{2})-3\mu A_{2}m_{2}(A_{1}^{2}+A_{2}^{2})+4A_{1}m_{2}c\omega\right]
\label{coeff_sol}%
\end{align}

which can be solved for $A_{1}\left(  \omega\right)  $, $A_{2}\left(
\omega\right)  $, $B_{1}\left(  \omega\right)  $, $B_{2}\left(  \omega\right)
$. Finally, the amplitude of the oscillation of inertial mass is calculated as%

\begin{equation}
\eta_{m,0}(\omega)=\sqrt{(A_{1}+B_{1}+x_{0})^{2}+(A_{2}+B_{2})^{2}}%
\end{equation}

\begin{figure}[ptbh]
\begin{center}
\centering\subfloat[\label{fig_app1}]{\includegraphics[height=54mm] {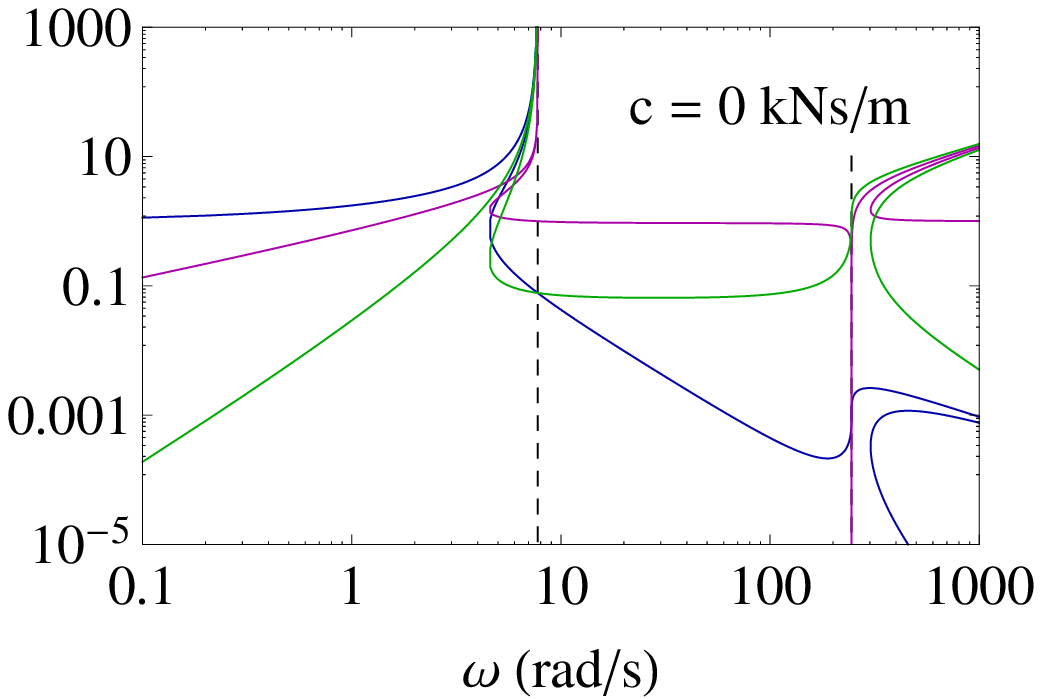}}
\hfill
\subfloat[\label{fig_app2}]{\includegraphics[height=54mm]{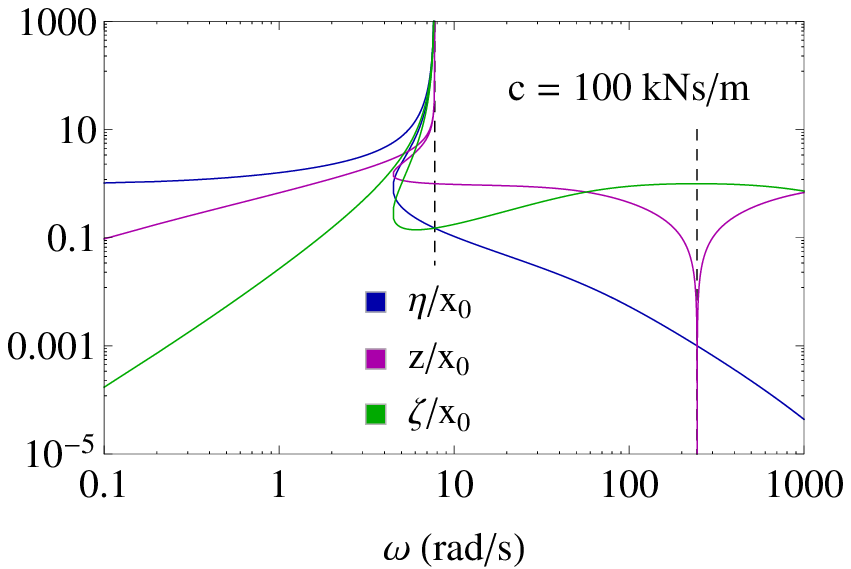}}
\end{center}
\caption{The frequency response of $\eta$, $z$, and $\zeta$ under undamped (a)
and damped conditions (b). Results are for $x_{0}=0.1$m. The two dashed
asymptotes are at $\omega_{1}=\sqrt{k_{s}/m_{2}}$ and $\omega_{2}=\sqrt
{k_{s}/m_{1}}$}%
\label{fig_app}%
\end{figure}

Figures \ref{fig_app} show the frequency response of the Duffing-like system
under different damping conditions.

\end{document}